\documentclass[useAMS,usenatbib]{mn2e}

\usepackage{amssymb}
\usepackage{amsfonts}
\usepackage{mncite}
\usepackage{epsfig}
\usepackage{psfig}
\usepackage{verbatim}

\newcommand{\asnow}{a_{\mathrm{snow}}}
\newcommand{\mjup}{M_{\mathrm{J}}}

\begin{document}

\title[How fast do Jupiters grow?]
{How fast do Jupiters grow? Signatures of the snowline and growth rate in the distribution
of gas giant planets}

\author[Ken Rice, Matthew T Penny \& Keith Horne]
 {Ken Rice$^1$\thanks{E-mail: wkmr@roe.ac.uk}, Matthew T Penny$^2$,  Keith Horne$^3$ \\
$^1$ SUPA\thanks{Scottish Universities Physics Alliance},
Institute for Astronomy, University of Edinburgh, Blackford Hill, Edinburgh, EH9 3HJ \\
$^2$ Department of Astronomy, Ohio State University, 140 W. 18$^{\rm th}$ Ave., Columbus, OH 43210 \\
$^3$ SUPA, School of Physics and Astronomy, University of St Andrews, North Haugh, St Andrews, Fife KY169SS}

\maketitle
\begin{abstract}
We present here	observational evidence that the snowline plays a significant role in the	
formation and evolution	of gas giant planets. When considering the	
population of observed exoplanets, we find a boundary in	
mass-semimajor axis space that suggests planets	are preferentially	
found beyond the snowline prior to undergoing gap-opening inward	
migration and associated gas accretion.	This is	consistent with	
theoretical models suggesting that sudden changes in opacity -- as	
would occur at the snowline -- can influence core migration. Furthermore,	
population synthesis modelling suggests that this boundary implies that
gas giant planets accrete $\sim 70$ \% of the inward flowing gas,	
allowing $\sim 30$ \% through to the inner disc.  	
This is qualitatively consistent with observations of transition discs suggesting the presence of
inner holes, despite there being ongoing gas accretion.	
\end{abstract}

\begin{keywords}
planets and satellites: formation --- Solar system: formation --- stars: pre-main-sequence --- planetary systems --- planetary systems: formation --- planetary systems: protoplanetary discs 
\end{keywords}
\section{Introduction}
Since the discovery, in 1995, of the first extrasolar planet around a main-sequence star \citep{mayor95}, 
a further 759 such planets have been detected.  Many of these planets have masses similar to that of Jupiter and hence are 
generally regarded as gas giants.  The standard model for the formation of these planets is the core accretion model \citep{pollack96}.   
In this model, micron-sized dust grains grow to form kilometre-sized planetesimals that then coagulate to form planetary mass 
bodies that, if sufficiently massive, may gravitationally attract a gaseous envelope if gas is still present in the disc.

Population synthesis models \citep{ida04,alibert05} have generally been successful in reproducing the properties of the observed 
exoplanet population.  However, while these models have illustrated how the overall process leads to a population consistent with 
that observed, they have not specifically quantified any individual parts of the process.

It has, however, recently been noted \citep{wright09} that the distribution in semimajor axis of the observed exoplanets shows a peak at $\sim 1$ AU.  
It has been suggested that this peak may be a consequence of disc dispersal through photoevaporation \citep{alexander12}. Some population 
synthesis models \citep{mordasini12} have also managed to reproduce this peak in the semimajor axis distribution and suggest that it is an 
imprint of the snowline. The ices that collect on small dust grains sublimate inside the snowline, where the temperature in the disc exceeds 170 K.  
The snowline radius is 2.7 AU for a solar-mass star \citep{hayashi81} and scales with $M^2$.
 
\begin{figure*}
\begin{center}
\psfig{figure=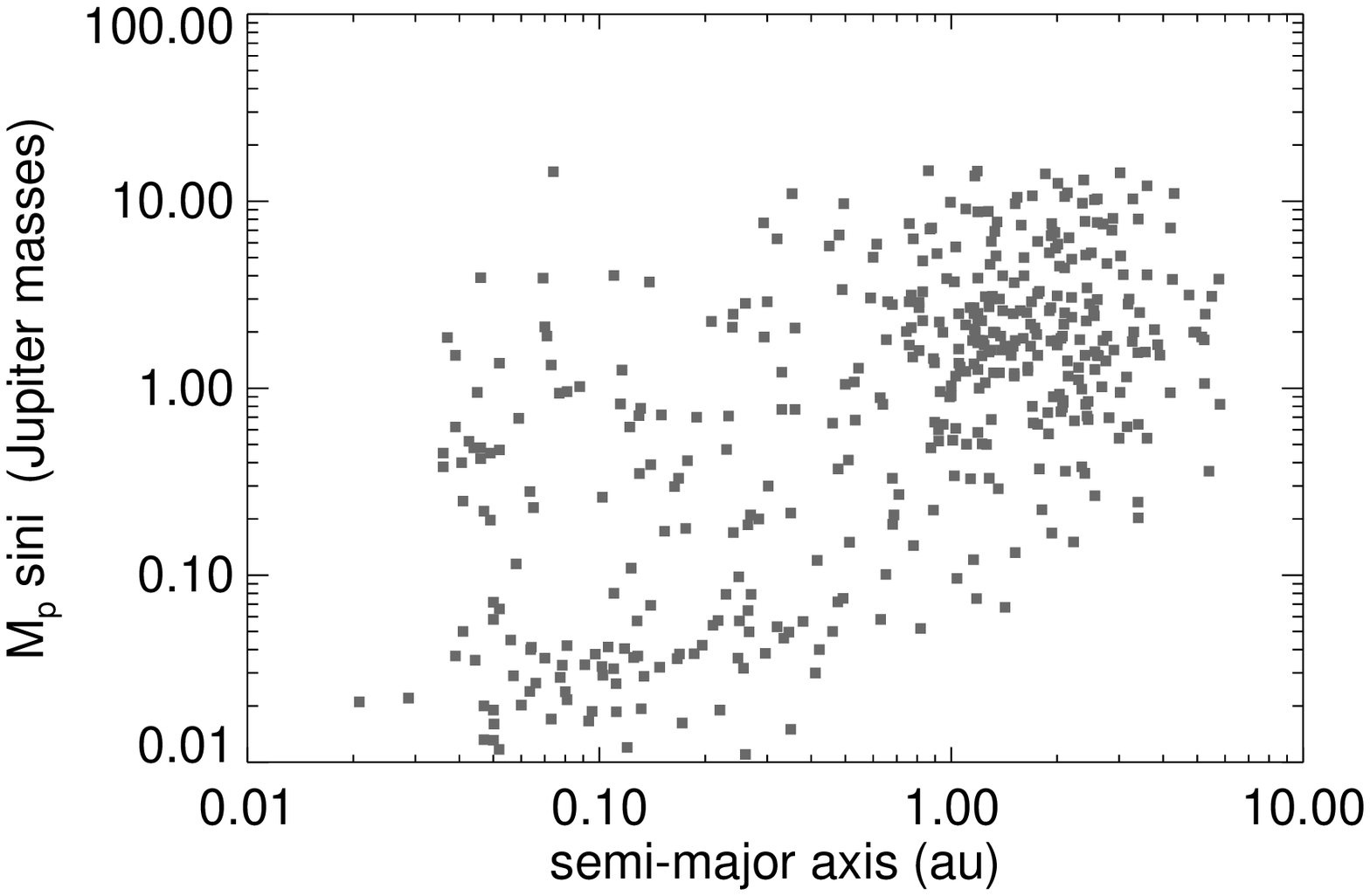,width=0.45\textwidth}
\psfig{figure=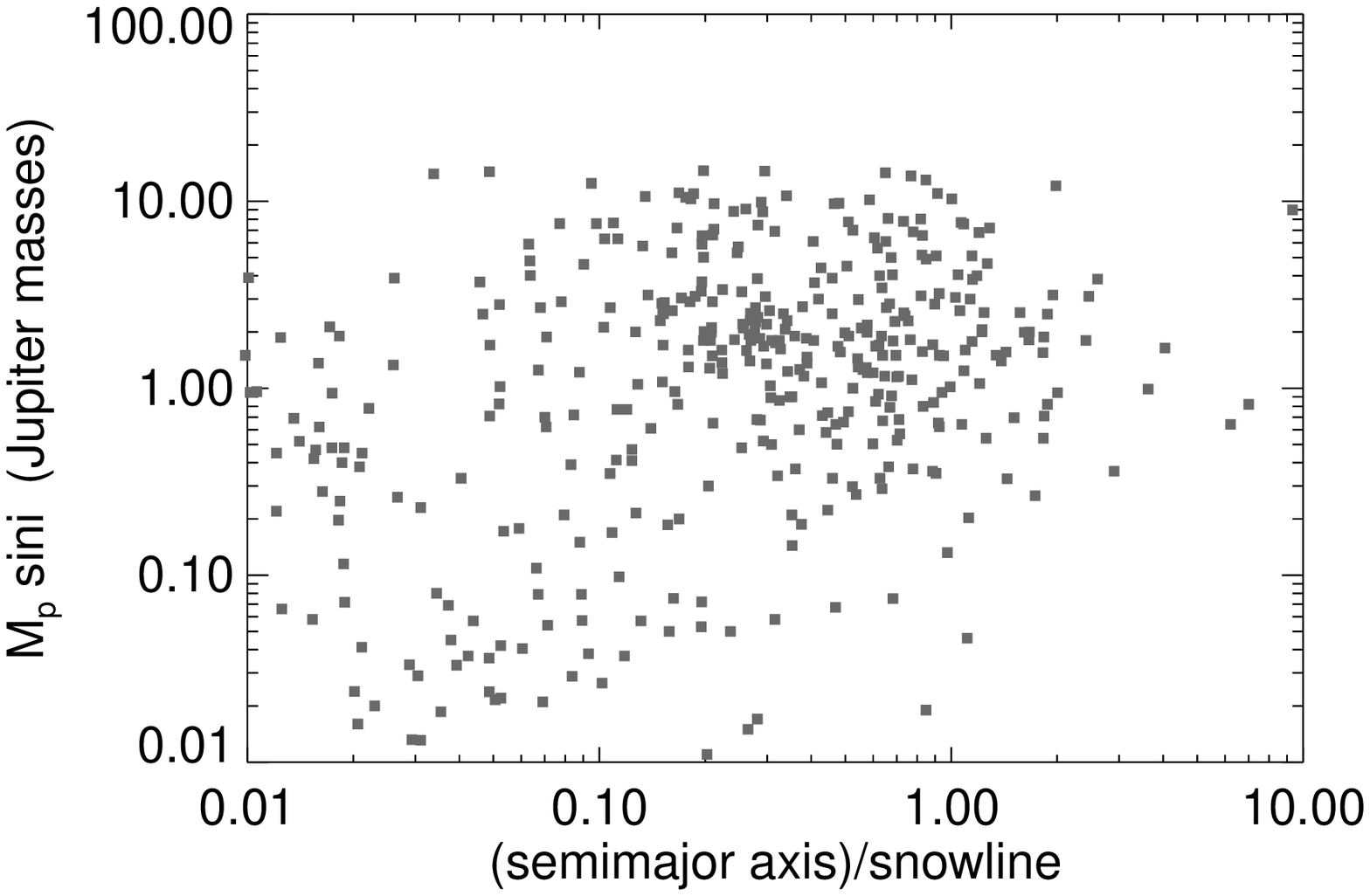,width=0.45\textwidth}
\caption{The left-hand panel shows planet mass against semimajor axis and illustrates an excess of planets beyond 1 AU \citep{wright09}. 
The right-hand panels shows planet mass against semimajor axis with the semimajor axis normalised
with respect to the snowline of each exoplanets host star. In the right-hand panel, there appears to be quite a well-defined diagonal boundary, 
across which there is a step increase in the density of planets.} 
\label{fig:planmass}
\end{center}
\end{figure*}

During the gas giant planet formation process, the rocky/icy core is expected to go through a phase of rapid inward migration, 
known as type I migration \citep{ward97}. Quite how these cores survive to form gas giant planets is still uncertain.  This migration may, however, be strongly affected by 
regions with a sudden change in opacity, as would occur at the snowline \citep{menou04}.  To investigate this we consider the distribution of planets
in semi-major axis space. In the left-hand panel of 
Figure \ref{fig:planmass} we plot planet mass against semimajor axis, while in the right-hand panel we plot planet mass against semimajor axis 
with the semimajor axis normalised with respect to the predicted snowline of the host star.  
We consider 463 exoplanets that were first detected by radial velocity measurements. In the left-hand panel there appears to be an excess of 
planets beyond 1 AU that has been highlighted by \citet{wright09} and others \citep{mordasini12,alexander12}.  In the right-hand panel, however,
there appears to be, for planet masses below a few Jupiter masses, quite a well-defined 
diagonal boundary across which there is a step increase in the density of planets.  

To characterise the diagonal boundary seen in the right-hand panel of Figure \ref{fig:planmass}, we assume that it has the form $M \sin i/M_{\rm Jup} = \delta \left(a/a_{\rm snow}\right)^\eta$.  We consider a box with $0.2 \le M/M_{\rm Jup} < 7$ and
$0.07 \le a/a_{\rm snow} < 1$ and determine the values of $\delta$ and $\eta$ that produce the biggest density contrast. 
The solid line shows the boundary that produces the largest density contrast while the dashed and dotted lines shows two other solutions that lie 
on the extremes of the bootstrapped 49\% confidence contour in $\delta - \eta$ space, discussed in more detail in Section \ref{sec:dataan}.

\begin{figure}
\begin{center}
\psfig{figure=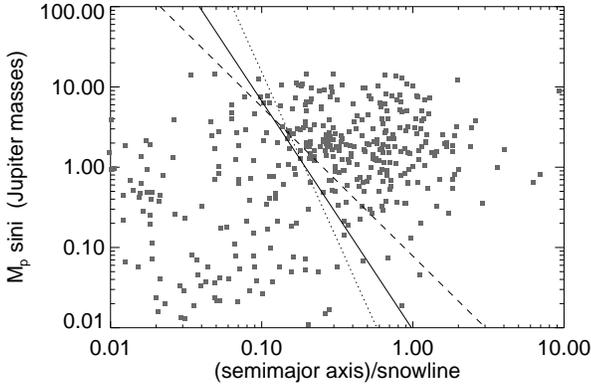,width=0.45\textwidth}
\caption{Planet mass against semimajor axis with the semimajor axis normalised 
with respect to the snowline of each exoplanets host star. There appears to be a diagonal boundary, in
mass-semimajor axis space, across which there is a step increase
in the density of planets. The solid line shows the boundary that
produces the the largest density contrast while the dashed and
dotted lines shows two other solutions that lie on the extremes of
the bootstrapped 49\% confidence contour in $\delta - \eta$ space.}
\label{fig:planmass_snowline}
\end{center}
\end{figure}

\begin{figure}
\begin{center}
\psfig{figure=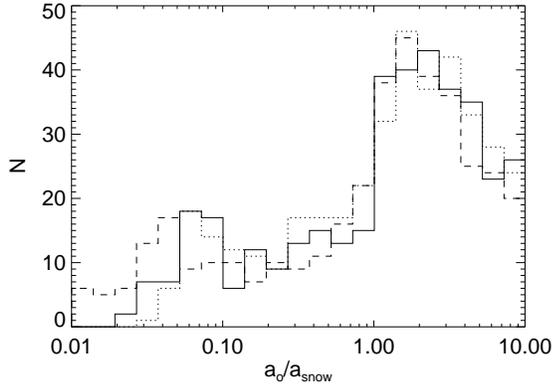,width=0.45\textwidth}
\caption{Distribution of the initial semimajor axis, $a_o$, of the exoplanet population with $a_o$ normalised with respect to the 
expected snowline of the host star.  In this case $a_o$ is determined by assuming that growth follows the diagonal line in 
Figure \ref{fig:planmass_snowline} and hence that $a_o = a (M_o/M_p)^{-1/\eta}$ with $M_o = \beta$ M$_{\rm Jup}$. The line styles 
correspond to those in Figure \ref{fig:planmass_snowline} and the largest density contrast occurs, as expected, for the best
fit line in Figure \ref{fig:planmass_snowline}.}
\label{fig:a_modsnowline_hist}
\end{center}
\end{figure}

If a planet is sufficiently massive, it can open a gap and migrate inwards via what is known as type II migration \citep{goldreich80,lin86}.  It can also continue growing by accreting 
some of the gas flowing though the gap \citep{lubow99}.  During this phase, planet growth typically follows a diagonal line in $\log M$ - $\log a$ space \citep{mordasini12}. 
We, therefore, assume that the diagonal boundary in the right-hand panel of Figure \ref{fig:planmass}, and characterised in Figure \ref{fig:planmass_snowline},
is a consequence of this migration and gas accretion. The diagonal lines
in Figure \ref{fig:planmass_snowline} have the form $M \sin i/M_{\rm Jup} = \delta \left(a/a_{\rm snow}\right)^\eta $ and so if $a_o = a_{\rm snow}$, 
$M_o = \delta$  M$_{\rm Jup}$, which we interpret as 
suggesting that a planet forming at the snowline reaches a mass of $\delta$ M$_{\rm Jup}$ before starting to migrate inwards via type II migration and 
growing via associated gas accretion.  For the best fit line (solid line), this corresponds to $M_o = 0.01$ M$_{\rm Jup}$, while for the dashed line 
it is $M_o = 0.078$ M$_{\rm Jup}$. These are both consistent with the mass at which we would expect type II migration
to start operating \citep{d'angelo03}.  The dotted line gives $M_o < 0.01$ M$_{\rm Jup}$ which may be unphysically small. We can, however, then 
use this to predict where 
each planet was prior to the start of this process.  If this process typically starts when the planets reach a mass of $M_o$, then from its 
current mass, $M_p$, and semimajor axis, $a$,  its initial semimajor axis would be $a_o = a (M_o/M_p)^{1/\eta}$.  We thus determine the 
initial distribution of these planets, as shown in Figure \ref{fig:a_modsnowline_hist}, where the initial semimajor axis is normalised to the snowline
and the line styles correspond to those in Figure \ref{fig:planmass_snowline}. 
Figure \ref{fig:a_modsnowline_hist} shows a definite jump at the snowline, with the largest density contrast occuring, as expected, for our
best fit line.  

The above suggests that prior to the final stages of planet formation 
(inward type II migration and final stage gas accretion), planets are preferentially found beyond the snowline of their host star.  
In this paper we, therefore, combine a realistic model of the evolution of protostellar discs with models of the growth and migration of gas 
giant planets to establish if this intepretation of the diagonal feature in Figure \ref{fig:planmass_snowline} is consistent with theoretical
models of planet migration and growth.
\section{Basic Model}
\subsection{Disc model}
\label{sec:disc}
We assume that the disc is axisymmetric and use the standard one-dimensional equations \citep{lyndenbell74, pringle81} to evolve
the surface density, $\Sigma(r,t)$. The surface density evolution is largely determined by the kinematic viscosity, $\nu$, and
by mass loss through a disc wind.  We assume that the viscosity has the form of an
$\alpha$ viscosity \citep{shakura73} so that $\nu = \alpha c_s H$, where $c_s$ is the disc sound speed, $H$ is the disc scaleheight, 
and $\alpha << 1$ is a parameter that determines the
efficiency of angular momentum transport.  It has recently been suggested \citep{owen10} that x-rays are the dominant driver
of photoevaporation, and so here we implement the x-ray photoionization model described in detail in \citet{owen11}. 

We ran 100 disc models and selected the central star mass randomly between $M_{\star} = 0.8$ M$_\odot$ and $M_{\star} = 1.2$ M$_\odot$.  We 
don't explicitly include the planets in these disc models, but instead use the range of disc models to later evolve planets with a large
range of different initial conditions.
We assume the disc extends from $r = 0.1$ AU to $r = 50$ AU with an initial surface density profile of $\Sigma \propto r^{-1}$.  In each
simulation the initial disc mass is $0.25 M_{\star}$. The initial disc mass is therefore quite high and such discs are likely
to be self-gravitating.  Within $50$ AU \citep{rafikov05}, it is expected that such discs will achieve a state of quasi-steady thermal equilibrium with 
dissipation due to the gravitational instability balanced by radiative cooling \citep{gammie01}. It has been shown \citep{balbus99,lodato04} that 
the gravitational instability then acts to transport angular momentum in a manner analogous to viscous transport. As described in detail in \citet{rice09} 
\citep[see also][]{clarke09,zhu09}, this can then be used to determine the effective value of $\alpha$ and, hence, the kinematic viscosity, $\nu$.

If, however, the effective gravitational $\alpha$ is less than $0.005$, we assume that another transport mechanism, such as the magnetorotational
instability (MRI) \citep{balus91} will then dominate and we set $\alpha = 0.005$. In addition, we also assume that irradiation from the 
central star sets a radially dependent minimum temperature in the disc \citep{hayashi81}.  Although all of our disc models start
with the same disc-to-star mass ratio, the variation in x-ray luminosity (from $5 \times 10^{28}$ erg s$^{-1}$ to $10^{31}$ erg s$^{-1}$)
produces a wide range of different disc lifetimes, largely consistent with that observed \citep{haisch01}.
We therefore have a set of disc models that can self-consistently evolve the surface density from the early stages, when the 
gravitational instability is likely to dominate, through to the later stages when an alternative transport mechanism, such
as MRI, will dominate and that also includes the late-stage dispersal due to photoevaporative mass-loss. 

\subsection{Core growth and migration}
Planet formation is thought to occur through the initial growth of micron-sized dust grains to, ultimately, kilometre-size planetesimals.  These 
planetesimals then continue to grow with, typically, the largest in any region dominating and undergoing what is known
as oligarchic growth \citep{kokubo98}.  If these oligarchs can grow sufficiently massive ($\sim 10$ M$_\oplus$) then, if there
is still gas present in the disc, they can gravitationally attract a gaseous envelope and can rapidly grow to 
become a gas giant planet.  

Once a planetary mass body has formed, it can exchange angular momentum with the surrounding disc material and migrate
radially \citep{goldreich80}.  There a number of different migration scenarios.  Low-mass planets ($M_{\rm pl} \sim 1$
M$_\oplus$) generate a linear disc response and migrate inwards through what is known as type I migration \citep{ward97}. 
High-mass planets are thought to open a gap in the disc gas and migrate through type II migration \citep{lin86}.  A
third type of migration, known as type III migration, may occur for intermediate mass planets \citep{masset03}. In the
type III regime, corotation torques generate very rapid migration which can be inward or outward \citep{masset03}.  

The migration rate for planets migrating in the type I regime increases with planet mass and analytic \citep{tanaka02}
and numerical \citep{bate03} estimates suggest that the timescale should be less than typical disc lifetimes.  These planets
should therefore migrate into the central star prior to becoming massive enough to enter the slower type II migration
regime. Recent numerical simulations
suggest that the type I migration rate can, however, be significantly reduced if
the disc thermodynamics is treated more realistically \citep{paardekooper08,paardekooper11}.  Recent population synthesis
models therefore typically assume that the analytic type I rate is reduced by at least a factor of about $30$
\citep{alibert05}. These models are now quite successful at reproducing the observed characteristics of the exoplanet
population \citep{mordasini09,alibert11}.  

Furthermore, it has been suggested \citep{menou04} that type I migration can be strongly influenced by sudden changes 
in the disc opacity, such as may occur near the snowline.  One of the goals of this paper is to investigate whether
or not the possible infuence of the snowline on core migration is consistent with the observed distribution of extraolar planets.  
After type I and, potentially, type III migration have 
distributed the planetary cores throughout the disc, they then undergo continued growth through gas accretion and
migrate via gap opening type II migration. We also want to compare the current distribution of exoplanets with
a population synthesis--like model to determine if we can quantify type II migration and the growth that accompanies
this migration process.

\subsection{Type II migration and gas accretion}
In our modelling we don't explicitly model the growth of the cores and the initial runaway gas accretion phase \citep{ikoma00,bryden00}
that occurs when the planetary core reaches the critical core mass \citep{ikoma00,papaloizou99}.  The runaway gas accretion phase
terminates when the planet reaches the gas isolation mass \citep{lissauer87}.  We assume that the gas feeding zone is 2 Hill radii wide with the Hill
radius of a planet with mass $M_p$ located at $a$ in the disc given by
\begin{equation}
r_H = a \left(\frac{M_p}{3 M_*} \right)^{1/3}.
\label{eq:Hill}
\end{equation}
One can then show that the isolation mass is 
\begin{equation}
M_{\rm iso} = \frac{(4 \pi a^2 \Sigma)^{3/2}}{(3 M_*)^{1/2}}.
\label{eq:iso}
\end{equation}

As discussed in section \ref{sec:disc} we have a series of disc models with central star masses between $M_* = 0.8$ M$_\odot$ and $M_* = 1.2$ M$_\odot$ 
and with a range of x-ray luminosities that results in a range of disc lifetimes that largely matches that observed \citep{haisch01}.  
The current exoplanet sample has more stars with mass $M_* > 1$ M$_\odot$ than with masses
$M_* < 1$ M$_\odot$ and so we use the disc models with central star masses $> 1$ M$_\odot$ twice as
often as those with central star masses $< 1$ M$_\odot$.

For each planet formation simulation, we assume that the planetary core forms and undergoes runaway gas accretion at a randomly chosen time between
$t = 1$ Myr and $t = 4$ Myr.  The disc model then gives us the mass of the central star and surface density and so Equation (\ref{eq:iso}) can be used to
determine the planet's isolation mass. The disc model also gives the time dependence of the disc viscosity, $\nu$. If sufficiently massive, a planet
will open a gap in the disc \citep{lin86}.  For a planet located at $a$, the width of the gap, $\Delta$, satisfies \citep{syer95}
\begin{equation}
\left(\frac{\Delta}{a}\right)^3 = \frac{\Omega a^2}{\nu} q^2,
\label{eq:width}
\end{equation}
where $q = M_p/M_*$. For a gap to open, the disc scaleheight, $H$, must be less than the gap width $\Delta$.  A second criterion is that the gap width needs to
be greater than the Roche radius, $R_L$, of the planet and hence
\begin{equation}
\Delta > R_L = q^{1/3} a \Rightarrow q > \frac{\nu}{\Omega a^2}.
\label{eq:Roche}
\end{equation}
If Equation (\ref{eq:Roche}) is satisfied and $H < \Delta$ then the planet is able to open a gap and migrate through type II migration.  

Type II migration effectively has two regimes, the ``disc dominated" regime \citep{armitage07} and the ``planet dominated" regime \citep{trilling98}.  
``Disc dominated" migration occurs when the local disc mass is large compared to the mass of the planet.  In this case the planet is 
coupled to the viscous evolution of the disc and the migration rate is independent of the mass of the planet.  The 
radial velocity of the planet is then
\begin{equation}
v_{pl} = - \frac{3 \nu}{2 a}.
\label{eq:discdommig}
\end{equation}
If, however, $M_p > 2 \Sigma a^2$, the inertia of the planet reduces its radial velocity to
\begin{equation}
v_{pl} = -\frac{3 \nu}{a} \frac{\Sigma a^2}{M_{p}}.
\label{eq:planetdommig}
\end{equation}
While migrating inwards, the planet is also able to accrete mass from the disc.  
For planets with mass of a few Jupiter masses, it has been suggested \citep{kley06} that the planet could accrete at a rate comparable
to gas accretion rate through the disc ($\dot{M_*} = 3 \pi \nu \Sigma$).  We therefore assume that in the ``planet dominated" migration regime the accretion rate onto the
planet is 
\begin{equation}
\dot{M}_p = \beta 3 \pi \nu \Sigma,
\label{eq:planaccr1}
\end{equation}
with $\beta$ typically close to unity (\citet{mordasini09} assume $\beta = 1$), and $\nu$ and $\Sigma$ coming from the self-consistent
disc models.  In the ``disc dominated" regime we assume that this is reduced by a factor $1/B$ where $B = (2 \pi \Sigma a^2)/M_p$
is essentially the ratio of the local disc mass to the planet mass.  The reduction is motivated by the mass accretion efficiency
for planet growth determined by \citet{veras04} based on the results of two-dimensional numerical simulations \citep{lubow99,d'angelo02}. 

We can combine Equations \ref{eq:planetdommig} and \ref{eq:planaccr1} to show that, in the ``planet dominated" regime 
$\log{M_p/M_o} = \log{\left(a/a_o \right)^{- \beta \pi}}$.  The dotted line in Figure \ref{fig:planmass_snowline} has a slope
steeper than $-\pi$ which, given that $\beta \leq 1$, would appear to be unphysical if our simple interpretation is correct. We almost certainly need more data
to determine, more accurately, the properties of this boundary.

\subsection{Putting it all together}
We randomly select one of our completed disc models and start modelling the evolution of the planet at the stage at which the 
planet semimajor axis has been set by type I (or type III) migration and the
planet has just undergone runaway gas accretion.  The initial planet
mass is given by Equation (\ref{eq:iso}) with $\Sigma$ taken from our chosen time-dependent disc model. Given the planet mass
and semimajor axis, we use Equations (\ref{eq:width}) and (\ref{eq:Roche}) to determine if the planet opens a gap in the disc.
If not, we assume that it grows slowly ($\dot{M}_{p} = 0.1 \dot{M}_*$ with $\dot{M}_*$ determined from our chosen disc model)
and remains at its initial semimajor axis until it either does satisfy the gap opening criteria, or the gas disc has dispersed. 
If the gap opening criteria are satisfied then Equation (\ref{eq:discdommig}) or (\ref{eq:planetdommig}) is used to determine
the inward migration rate. The rate at which the planet grows is then given by Equation (\ref{eq:planaccr1}) with $\beta$ either
a constant (in the ``planet dominated" regime) or a constant reduced by $1/B$ (in the ``disc dominated" regime).
We stop either when the planet reaches $a = 0.1$ AU (the inner edge of our disc models) or when the gas disc has
dispersed. To synthesise a population of exoplanets, we repeat the above for different disc models and for different initial starting times
and initial semimajor axes.  

\section{Results}
In these simulations we assume that the sudden change in opacity at the snowline \citep{menou04} results in planetary 
cores being preferentially found beyond the snowline, rather than inside the snowline.  We therefore assume, somewhat arbitrarily, that the 
exoplanet density is about 6 times lower inside the snowline than it is just outside the snowline.  We also 
assume that type I migration results in a pile-up at the snowline and hence that the distribution beyond the 
snowline is uniform in $\log a$.  The only other free parameter we have is $\beta$, the ratio of the gas accretion rate 
onto the planet to the gas accretion rate through the disc.

\begin{figure*}
\begin{center}
\psfig{figure=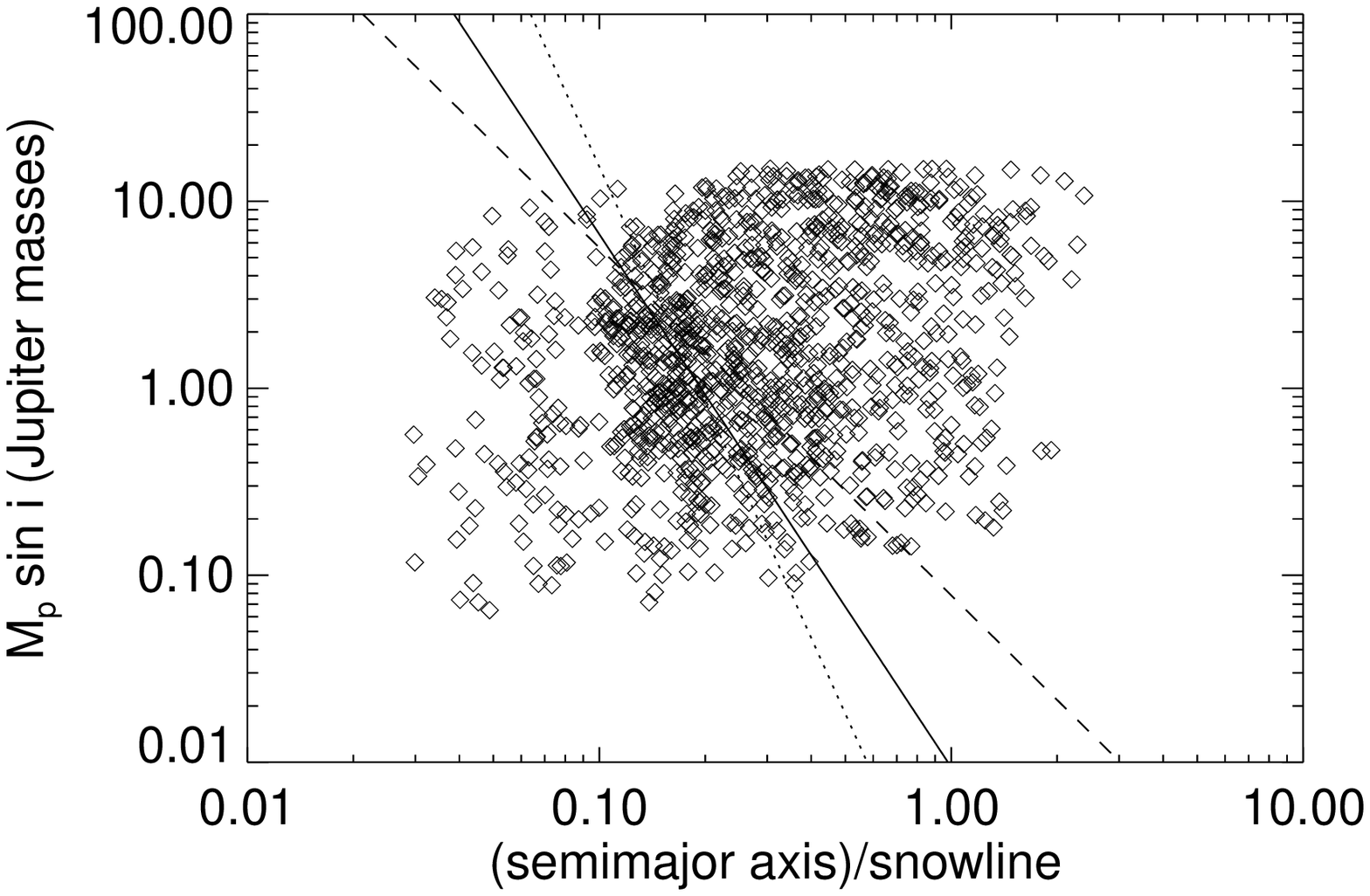, width=0.31\textwidth}
\psfig{figure=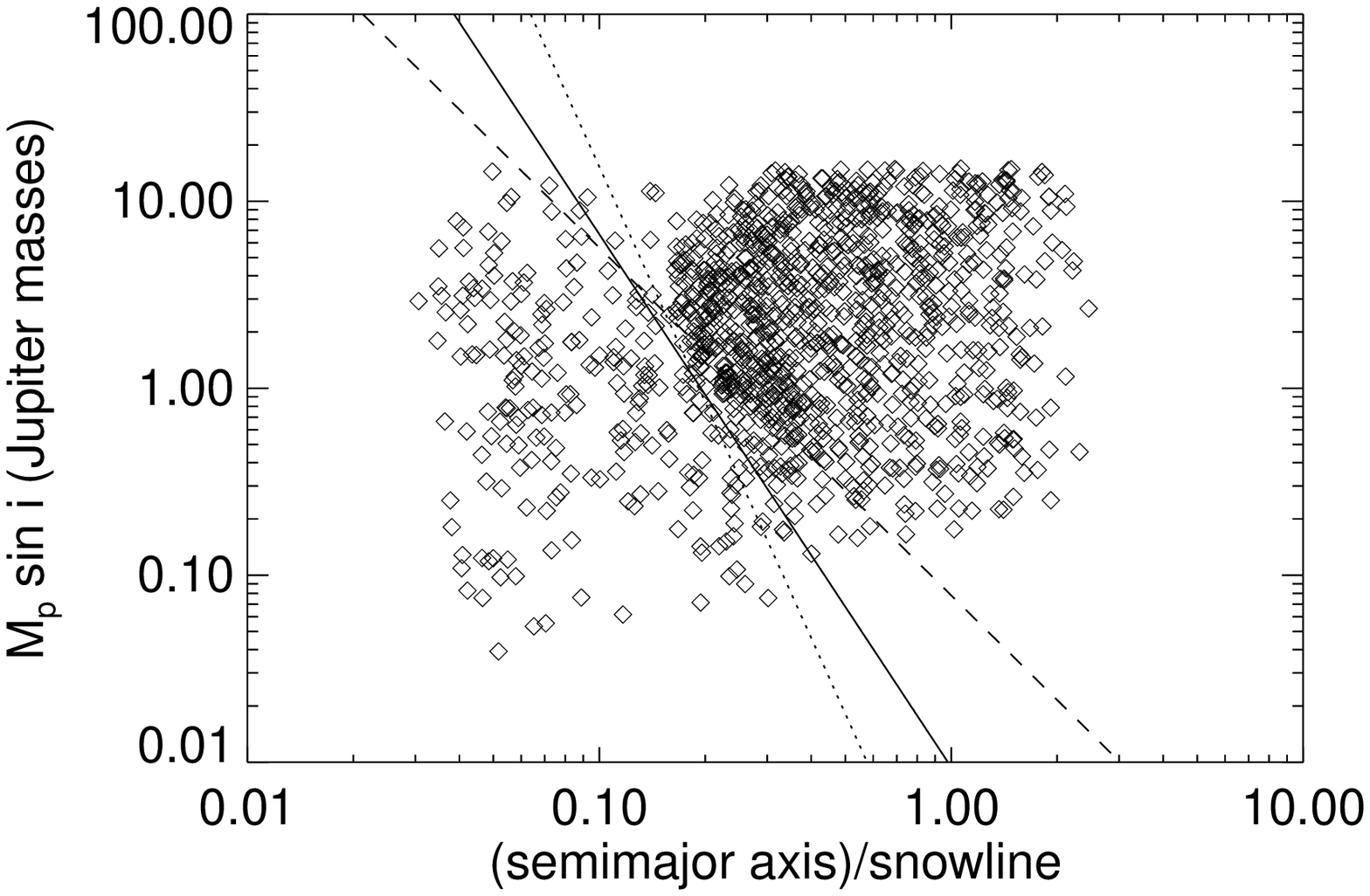, width=0.31\textwidth}
\psfig{figure=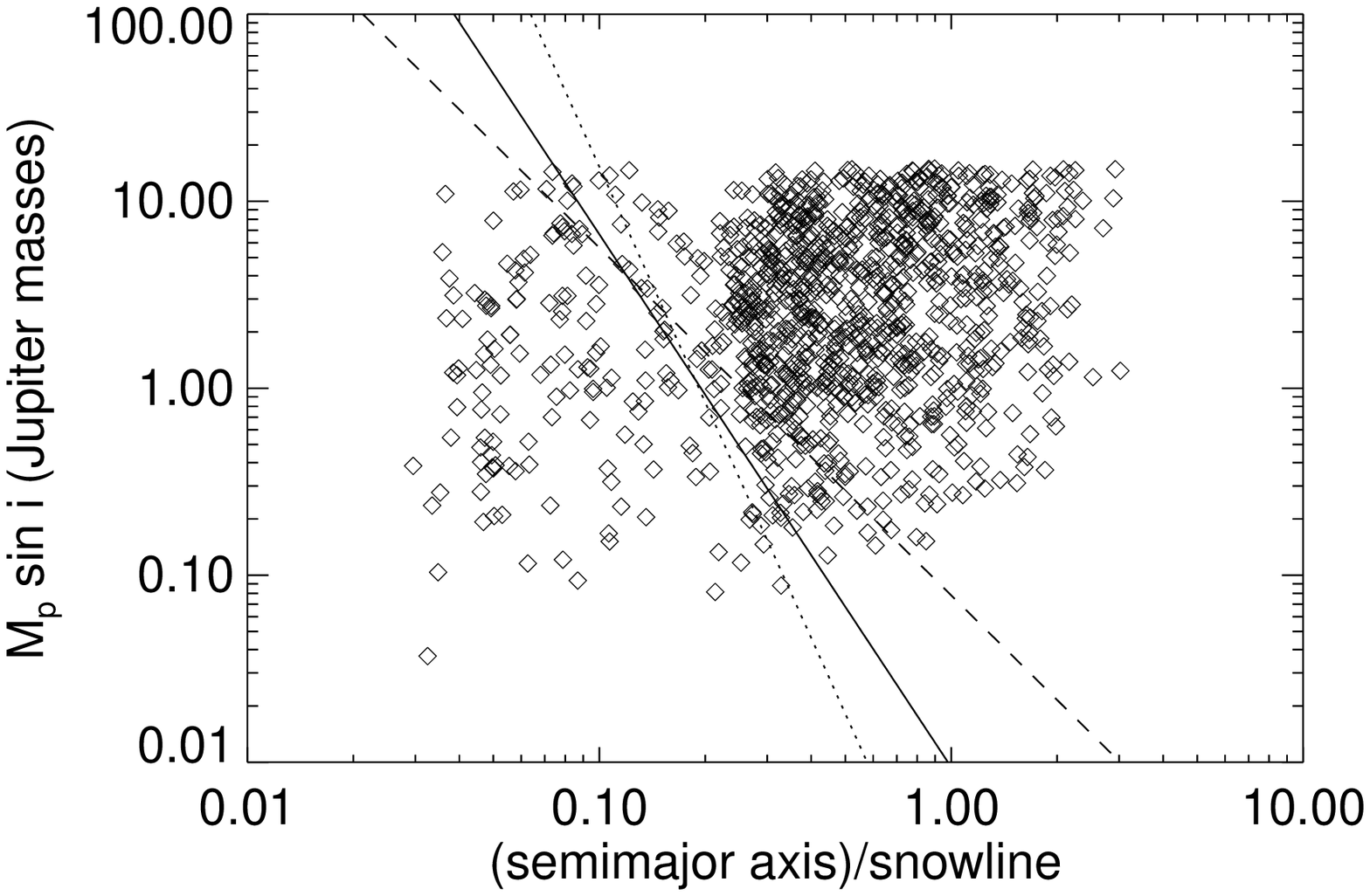, width=0.31\textwidth} \\
\psfig{figure=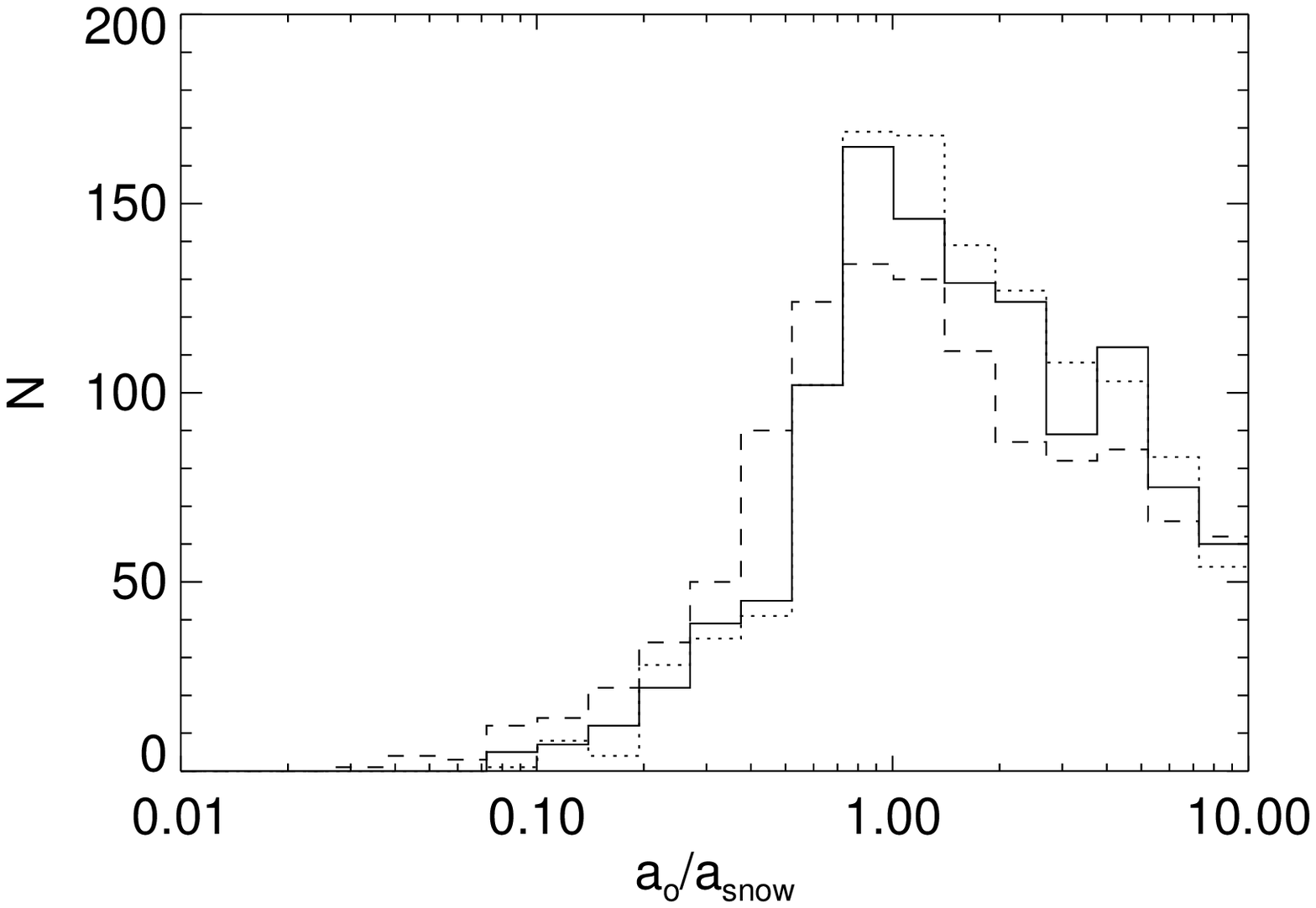, width=0.31\textwidth}
\psfig{figure=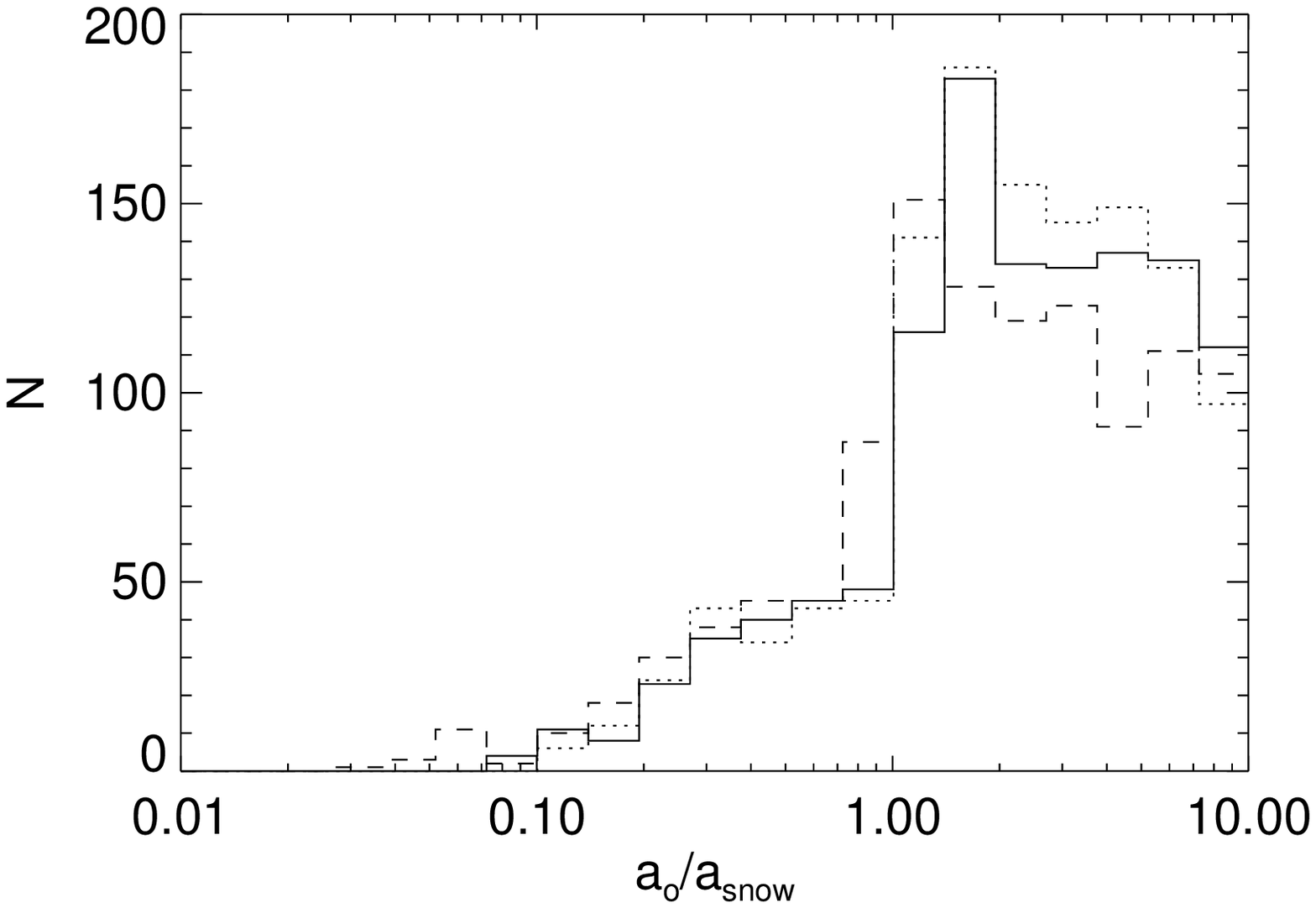, width=0.31\textwidth}
\psfig{figure=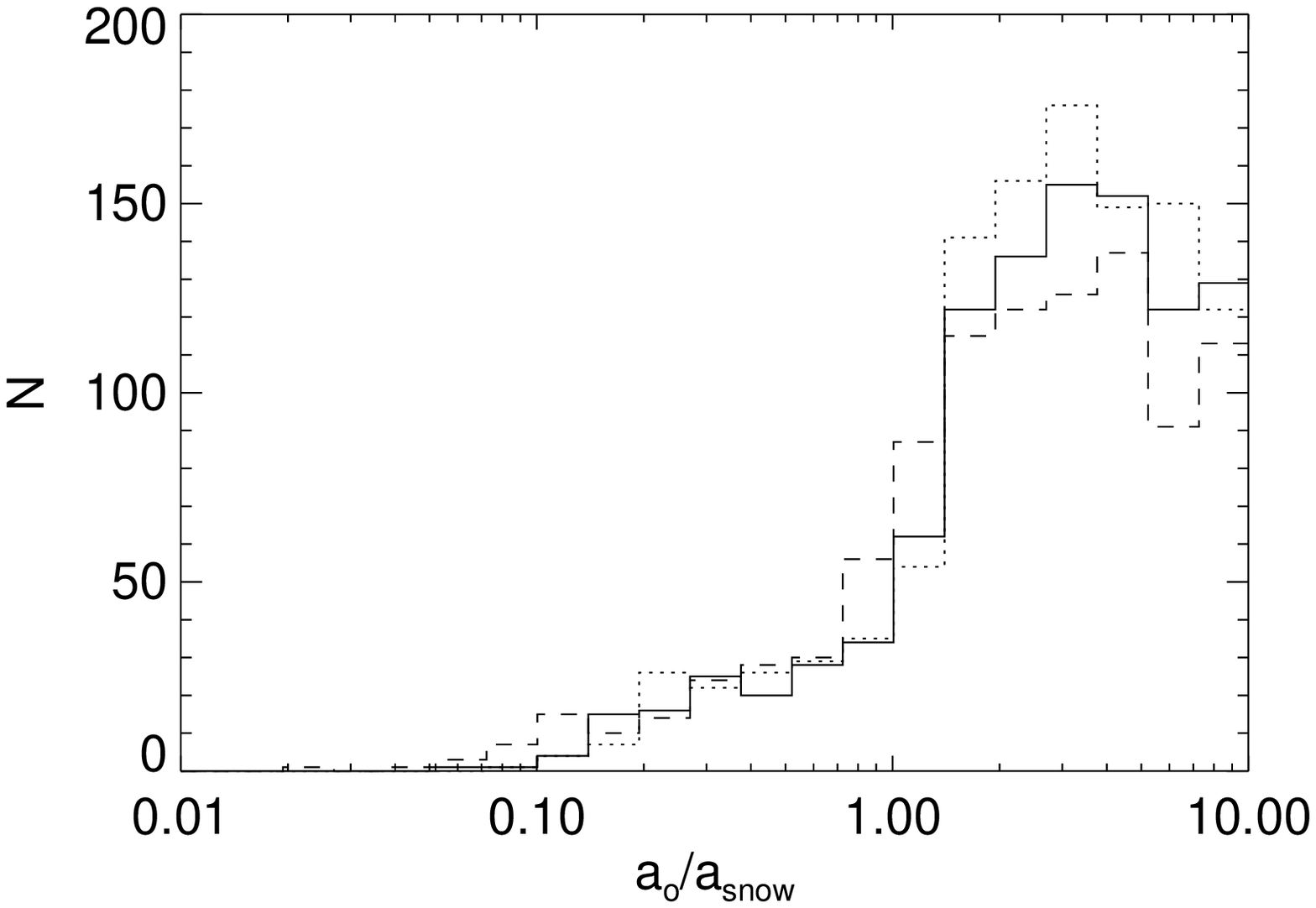, width=0.31\textwidth} \\
\psfig{figure=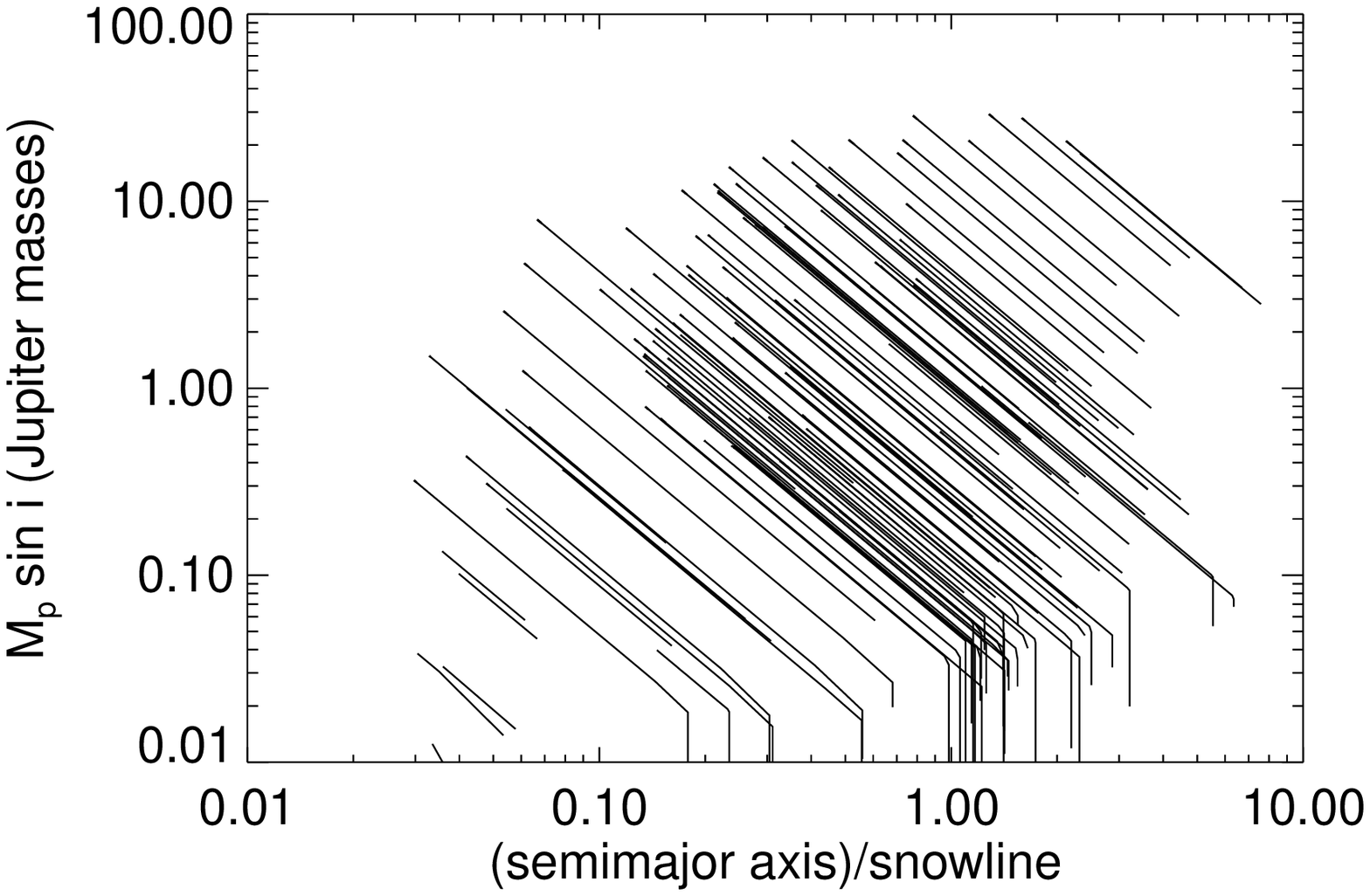, width=0.31\textwidth}
\psfig{figure=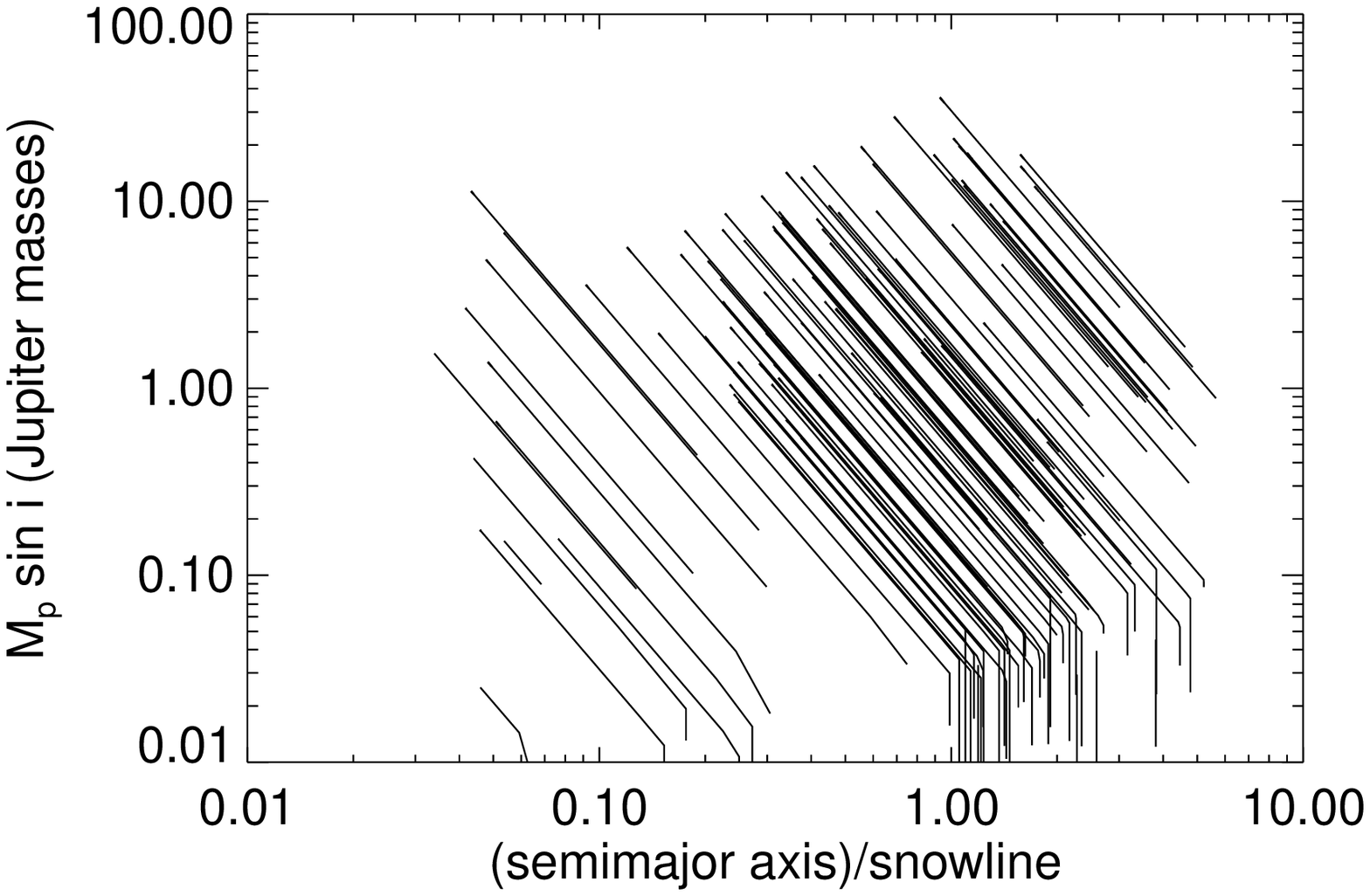, width=0.31\textwidth}
\psfig{figure=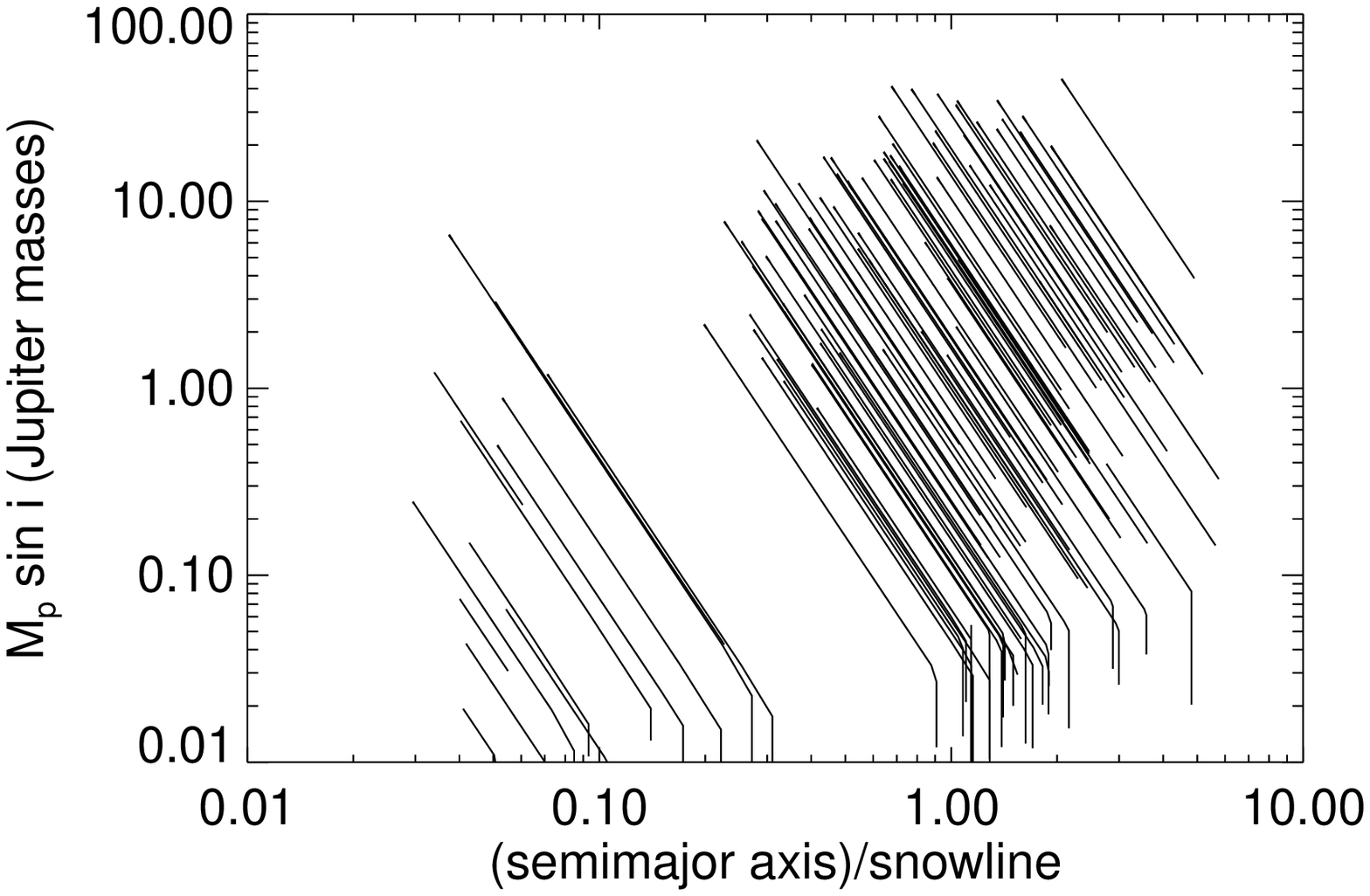, width=0.31\textwidth}
\caption{Series of figures for 3 different planet accretion rates.  The planetary accretion is represented by $\beta$, 
the ratio of the accretion rate onto the planet to the gas accretion rate through the disc.  The left-hand panel is for
$\beta = 0.5$, the middle panel is for $\beta = 0.7$, and the right-hand panel is for $\beta = 0.9$. The top row shows 
the final state distribution of planet mass and orbit semimajor axis, normalised with respect to the snowline, after the 
planet’s upward growth and inward migration are halted by photo-evaporation of the disc.  The middle panel is the distribution 
of the initial semimajor axis ($a_o$ – determined in the same way as for Figure \ref{fig:a_modsnowline_hist} and normalised with
respect to the snowline of the host star) for the 3 different boundaries
illustrated by the solid, dashed and dotted lines in the top set of figures. The bottom is a sample of $100$ planet growth tracks showing how the 
planets grow while migrating inwards through the disc.}
\label{fig:compilation}
\end{center}
\end{figure*}

Figure \ref{fig:compilation} shows results for 3 different planet growth rates, $\beta$. The left-hand panel is for $\beta = 0.5$, 
the middle panel is for $\beta = 0.7$, and the right-hand panel is for $\beta = 0.9$.  The top row of figures shows 
planet mass against semimajor axis for $\sim 1000$ simulated planets and with the semimajor axis normalised with respect to the 
snowline of the planet's host star.  For each planet, we have also randomly selected an inclination angle such as to produce an 
isotropic distribution of orbit orientations. We also only show those that would have been detected with a $20$ year radial velocity 
campaign with a cadence of $1$ month and a rms velocity sensitivity of $1$ m s$^{-1}$. The diagonal lines are the same as 
that in Figure \ref{fig:planmass_snowline}.  Figure \ref{fig:compilation} already shows that for $\beta = 0.5$ the density jump occurs
inside the diagonal lines, while for $\beta = 0.9$, it appears to be - largely - beyond the diagonal lines. The best match to that observed 
(Figure \ref{fig:planmass_snowline}) appears to be the middle figure in the top panel in Figure \ref{fig:compilation} in which $\beta = 0.7$. 

As discussed in relation to Figure \ref{fig:a_modsnowline_hist}, we can use the diagonal lines to predict where each planet 
emerged prior to the start of type II migration and associated gas accretion.  This is shown in the middle row of Figure \ref{fig:compilation}. The middle figure of this row 
($\beta = 0.7$) is very similar to that seen in Figure \ref{fig:a_modsnowline_hist} and suggests that, prior to type II migration, 
the planet population peaks at the snowline and that planets then grow in a manner consistent with the diagonal boundary in 
Figure \ref{fig:planmass_snowline}.  However, for $\beta = 0.5$ and $\beta = 0.9$, the peaks fall inside and beyond the snowline respectively.   

The slope of the diagonal boundary is largely determined by the relationship between the rate at which the planet migrates inwards and the 
rate at which it grows.  The bottom row of Figure \ref{fig:compilation} shows a sample of growth tracks for each $\beta$ value. 
This illustrates that the slope of the growth tracks in mass-semimajor axis space depends on the value of $\beta$ 
(the ratio of the accretion rate onto the planet to the gas accretion rate through the disc).  The slope of the boundary in mass-semimajor axis
space (top figures in Figure \ref{fig:compilation}) appears not to precisely match the slope of the growth tracks, but is clearly influenced by
the slope of these growth tracks.  If $\beta > 0.7$, 
the diagonal line would be steeper than that observed in Figure 1, while if $\beta < 0.7$, the diagonal line would be shallower.
The tracks also show that these planets are typically migrating in the ``planet dominated" regime rather than in the ``disc dominated" regime.
\section{Discussion and Conclusions}
We use self-consistent disc simulations together with models of gas giant planet migration and growth to show that if the snowline 
influences core migration, as suggested by \citet{menou04}, we can largely reproduce the observed planet mass against semimajor axis distribution.  
This in itself is interesting as it indicates that the snowline plays a crucial role in preventing planetary cores from migrating 
(via type I migration) into the host star.  Furthermore, we use this to quantify the gas accretion rate onto the planet during the 
“planet dominated” type II migration phase.  We find that it must accrete at a rate of about $70$ \% that of the gas accretion rate through the disc.  
This is the first time that we have been able to quantify the rate at which a gas giant planet grows during the final stages of disc evolution.  

Our results indicate that during type II migration a gas giant planet consumes $\sim 70$ \% and allows $\sim 30$ \% of the gas to flow inward. As a result, 
we expect a slightly lower gas density inside the planet’s orbit.  It has also been suggested, however, that dust filtration at the gap edge \citep{rice06} will
prevent all but the smallest dust grains from reaching the inner disc, significantly enhancing the gas-to-dust ratio.  This is consistent with observations of transition discs \citep{espaillat10} which have near-IR deficits indicating a lack of warm dust in the inner disc, but still appear to be accreting at TTauri--like rates.  
\section*{acknowledgements}
W.K.M.R. acknowledges support from the Scottish Universities Physics Alliance (SUPA) 
and for support from the Science and Technology Facilities Council (STFC) through
grant ST/J001422/1.

\appendix

\section{Data analysis}
\label{sec:dataan}

To measure the slope of the observed linear feature in $\log{M/M_J}$-$\log{a/a_{\rm snow}}$ space,
we aim to fit a straight line
\begin{equation}
\log(a/\asnow) = c + \alpha[\log(M \sin i/\mjup)-\log 7],
\label{fittingline}
\end{equation}
to the data so that it maximizes the density ratio on either side of the line.  This is related to the fitting parameters ($\delta$,$\eta$) that we use in the paper above through
$\eta = 1/ \alpha$ and $\delta=10^{c - \alpha \log 7}$.
To exclude other features of the simulated and observed planet distributions we select only planets with $0.2\le M/\mjup<7$ and $0.07\le a/\asnow<1$, and measure the planet density on each side of this box, as divided by the line. In order to penalize solutions that maximized the density ratio by producing a small area on either side of the line, we actually maximized the quantity $\sqrt{N_{+}N_{-}}\rho_{+}/\rho_{-}$, which we call the fitting metric, where $\rho_{+/-}$ is the planet density to the right/left of the line, and $N_{+/-}$ is the number of planets to the right/left of the line. For the observed planets, to account for the varying detection efficiency as a function of the radial velocity semiamplitude $K$, we considered planets with $K<2.84$~m~s$^{-1}$ to contribute a count of $(2.84$~m~s$^{-1}/K)$ to $\rho_{+/-}$ and $N_{+/-}$.

\begin{figure}
\psfig{figure=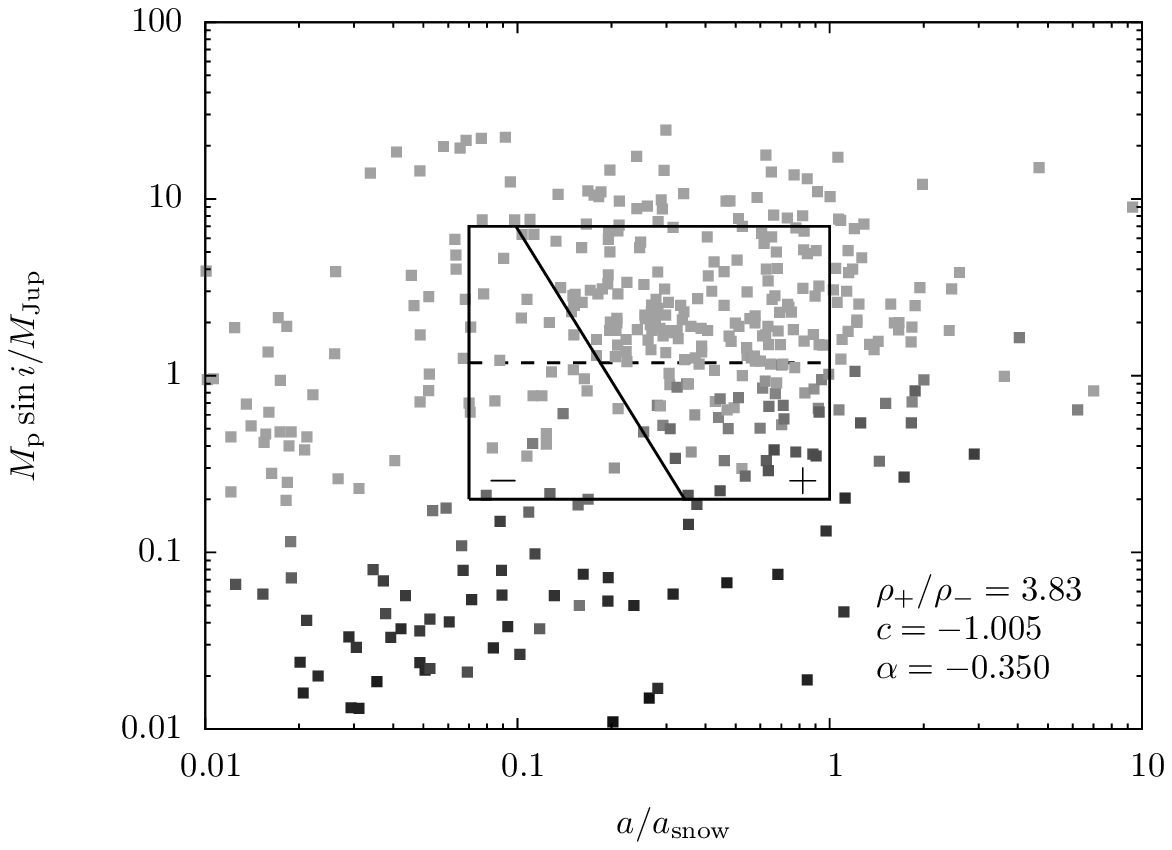, width=0.45\textwidth}
\caption{Plot of $M \sin i$ against $a/\asnow$ for the observed planet distribution. The box shows the selection of planets that are used to fit a linear density feature. The diagonal line (equation~\ref{fittingline}) is the best fitting line that maximizes the density ratio on each side of the line. Other details are explained in the text.}
\label{fig:fittedline}
\end{figure}

Figure~\ref{fig:fittedline} shows the observed planet distribution with the best fitting line, which has parameters $c=-1.005$ and $\alpha=-0.350$. It is clear by eye that the number density of planets is significantly higher to the right of this line -- the density contrast is $\rho_{+}/\rho_{-}=3.83$. In the plot, planets are shaded by their RV semiamplitude, such that the planets with the lightest grey are unweighted (i.e. $K>2.84$~m~s$^{-1}$ and they count as $1$ planet) and black indicates $K=0$~m~s$^{-1}$. The point with the smallest RV semiamplitude in the fitting box has $K=0.91$~m~s$^{-1}$ and so counts as $3.1$ planets.

\begin{figure}
\psfig{figure=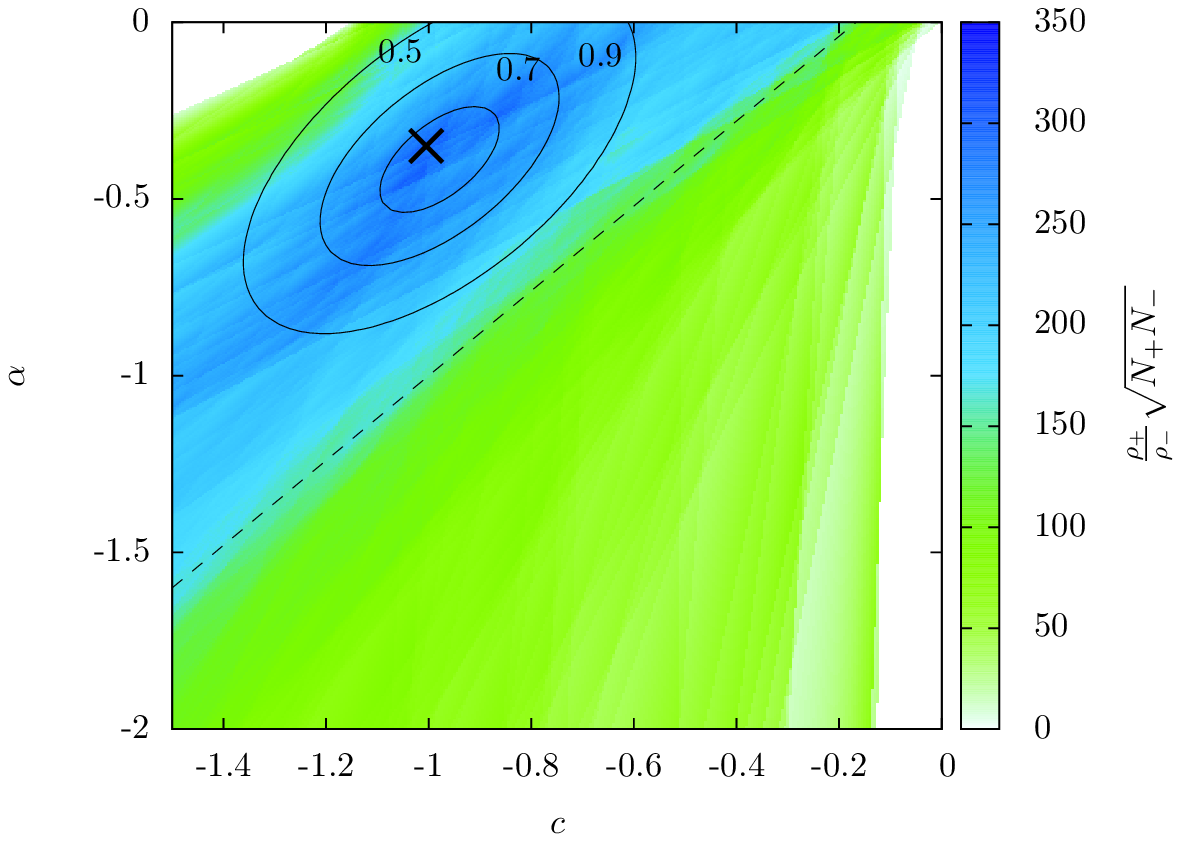, width=0.45\textwidth}
\caption{Map of the fitting metric plotted against the parameters $c$ (intercept) and $\alpha$ (slope) of the line fitted to the density feature seen in the observed planet distribution. The cross marks the best-fit solution for the observed distribution, and numbers denoting the $\beta$ value for simulated planet distributions are placed at the corresponding best-fit parameters. The black contours show estimated $68$- and $95$-percent confidence intervals for the observed planet solution based on a form of bootstrap resampling, while the dashed line was used to discard outlier solutions from this bootstrap analysis.}
\label{fig:paramspace}
\end{figure}

Figure~\ref{fig:paramspace} shows the fitting metric $\sqrt{N_{+}N_{-}}\rho_{+}/\rho_{-}$ mapped against the two parameters of the line. However, it is not possible to estimate the parameter uncertainties from the fitting metric as it does not obey $\chi^2$ statistics. Instead, we can estimate the uncertainties by a form of bootstrapping.  To estimate the uncertainties by bootstrapping, we produced fake planet distributions by Poisson sampling from the observed planet number density (including $K$ weighting) in four uneven quadrants of the fitting box. The interior boundaries of the quadrants were defined by the best fitting line and the horizontal dashed line in Figure~\ref{fig:fittedline} at the center of the selected mass range. The density in each quadrant was assumed to be uniform in $\log(a/\asnow)$ and $\log(M \sin i/\mjup)$. Each fake distribution is then fitted in the same way as the observed distribution, yielding an estimate of the parameters $c$ and $\alpha$ in each case. A small number ($6$~percent) of the bootstrap samplings produce solutions far from the best fit parameters for the observed planets, and are distributed roughly uniformly over the lower right half of the $c$-$\alpha$ parameter space (in the lower, green region in Figure~\ref{fig:paramspace}), while the majority of samplings lie near the observed planet solution, with a distribution that roughly follows the degeneracy suggested by the map of the fitting metric. The outlier points significantly skew estimates of the error ellipse so we discard samples that fall below the dashed line in Figure~\ref{fig:paramspace}. The contours in the figure show the $49$-, $70$- and $85$-percent confidence intervals for a Gaussian with mean and covariances matching the remaining bootstrap sample, while the cross marks the position of the solution for the observed planet distribution and numbers indicate the solutions for the simulated data sets with different values of $\beta$. It is clear that of the simulated planet distributions, the properties of the linear density feature in the simulation with $\beta=0.7$ best match those of the feature in the observed distribution.

\end{document}